\input epsf
\documentstyle{mn}
\newif\ifAMStwofonts
\def\lesssim{\mathrel{\hbox{\rlap{\hbox{\lower4pt\hbox{$\sim$}}}\hbox{$<$}}}}
\def\gtrsim{\mathrel{\hbox{\rlap{\hbox{\lower4pt\hbox{$\sim$}}}\hbox{$>$}}}}
\def\apj{ApJ}

\def\aj{AJ}
\def\aap{A\&\hskip-1pt A}

\def\mnras{MNRAS}
\def\pasp{PASP}
\def\araa{ARA\&\hskip-1pt A}
\def\nat{Nature}
\def\pasj{PASJ}

\newcommand{\rvec}  {\mbox{\boldmath $r$}}

\title[Microlensing Faint Binary Companion Detection]
      {Another Channel to detect Close-in Binary\\
       Companions via Gravitational Microlensing}
\author[Chang \& Han]
       {Heon-Young Chang$^\dagger$ \& Cheongho Han$\ddagger$\\
         ${}^\dagger$ Korea Institute for Advanced Study, \\
         207-43 Cheongryangri-dong Dongdaemun-gu, Seoul 130-012, Korea\\
	{\tt hyc@ns.kias.re.kr}\\
	       \\
        ${}^\ddagger$ Dept.\ of Physics, \\
        Chungbuk National University, Chongju 361-763, Korea\\
        {\tt cheongho@astroph.chungbuk.ac.kr}}

\date{Accepted:\\
      Received: }
\pagerange{\pageref{firstpage}--\pageref{lastpage}}
\begin{document}

\maketitle
\label{firstpage}

%==== Abstract ==========================================================
\begin{abstract}
Gaudi \& Gould (1997) showed that close companions of remote binary systems 
can be efficiently detected by using gravitational microlensing via the 
deviations in the lensing light curves induced by the existence of the lens 
companions.  In this paper, we introduce another channel to detect faint 
close-in binary companions by using microlensing.  This method utilizes a 
caustic-crossing binary lens event with a source also composed of binary 
stars, where the companion is a faint star.  Detection of the companion is 
possible because the flux of the companion can be highly amplified when it 
crosses the lens caustic.  The detection is facilitated since the companion 
is more amplified than the primary because it, in general, has a smaller 
size than the primary, and thus experiences less finite source effect. 
The method is extension of the previous one suggested to detect close-in 
giant planets by Graff \& Gaudi (2000) and Lewis \& Ibata (2000) and further 
developed by Ashton \& Lewis (2001).  From the simulations of realistic 
Galactic bulge events, we find that companions of K-type main sequence or 
brighter can be efficiently detected from the current type microlensing 
followup observations by using the proposed method.  We also find that 
compared to the method of detecting lens companions for which the efficiency 
drops significantly for binaries with separations $\lesssim 0.2$ of the 
angular Einstein ring radius, $\theta_{\rm E}$, the proposed method has 
an important advantage of being able to detect companions with substantially 
smaller separations down to $\sim ({\cal O}) 10^{-2}\theta_{\rm E}$.
\end{abstract}

%=======================================================================

\section{Introduction}
Observationally constraining the characteristics of faint close-in binary 
companions is important to understand the process of star formation and 
evolution.  For example, mass transfer between the binary components 
through Roche lobe overflow could alter the physical processes of stellar 
evolution (Iben \& Tutukov 1984; Han, Tout, \& Eggleton 2000).  Theoretical 
investigations about the origin of Type Ia supernova are also based on the 
observations of close binaries (Nomoto \& Sugimoto 1977; Li \& van den Heuvel 
1997; Langer et al.\ 2000).  Another possible outcome of close binary 
evolution is a merger during the late stage of a common envelope phase 
(Iben \& Livio 1993; Armitage \& Livio 2000; Taam \& Sandquist 2000),
whose physical parameters should be constrained by observations.

Detection of faint close-in binary companions is also important for the 
accurate determination of the stellar mass function.  The mass function of 
stars is derived from their luminosity function by using the mass-luminosity 
relation.  In this determination, however, if the mass function is derived 
without considering unresolved faint binary companions, the resulting 
stellar mass function will be biased towards larger masses.  Even if they 
are taken into consideration, the constructed mass function will suffer 
from large uncertainties because the uncertainties in the properties (the 
frequency and the mass ratio distribution) of the unresolved binaries will 
directly propagate into the resulting mass function (Kroupa, Tout, \& 
Gilmore 1991; Reid 1991; Reid \& Gizis 1997).

Faint close-in binary companions can be detected by using various methods, 
e.g.\ direct imaging by using high-resolution interferometers, astrometric 
measurements of the primary motion, spectroscopic radial velocity 
measurements, and photometric measurements of the light variation caused by 
occultation (eclipsing).  However, these methods have limitations in 
detecting and characterizing various types of binaries.  For example, the 
direct imaging and astrometric methods have limited applicabilities only 
to nearby stars.  The spectroscopic method can be applied to detect binaries 
located at large distances.  However, determining the mass ratio between 
the binary components by using this method requires that both components 
should be bright for the construction of the individual stars' radial 
velocity curves, limiting the applicability of the method.  Constraining 
the physical properties of eclipsing binary companions suffers from similar 
limitation.  Therefore, invention of complementary methods to detect and 
characterize various types of binaries is important.

Close binaries can also be detected and characterized by using microlensing.
Detecting binary companions via microlensing is possible because if events 
are caused by lenses composed of binaries, the resulting light curves can 
exhibit noticeable deviations from the smooth and symmetric ones of single 
lens events.  In addition, one can constrain the physical properties of the 
binaries (the mass ratios and the scaled projected separation between the 
binary components) by fitting model light curves to the observed one.

In this paper, we introduce another method of detecting and characterizing 
faint close-in binary companions by using microlensing.  This method utilizes 
caustic-crossing binary lens events with a source also composed of binary 
stars, where the companion is a faint star.  Detecting the source companion 
is possible because the flux from the companion can be highly amplified 
when it crosses the lens caustics.  This method is extension of the previous 
one suggested to detect close-in giant planets by Graff \& Gaudi (2000) and 
Lewis \& Ibata (2000) and further developed by Ashton \& Lewis (2001).

The paper is organized in the following way.  We describe the basics of 
microlensing in \S\ 2.  We perform simulations of example Galactic bulge 
events in \S\ 3 and evaluate the detectability of the deviations induced 
by binary source companions for the example events in \S\ 4.  A brief 
summary of the results and discussion about the advantages and disadvantages 
of the proposed method compared to that of detecting lens companions are 
included in \S\ 5.

\section{Basics of Microlensing}

The light curve of a single point-source event caused by a single point-mass 
lens is represented by a simple analytic form of
\begin{equation}
A_0 = {u^2+2\over u\sqrt{u^2+4}},
\end{equation}
where $u$ is the lens-source separation normalized by the angular Einstein 
ring radius.  The Einstein ring radius is related to the physical parameters 
of the lens by
\begin{equation}
\theta_{\rm E} = \sqrt{4GM \over c^2}
       \left( {1\over D_{\rm ol}} - {1\over D_{\rm os}} \right)^{1/2},
\end{equation}
where $M$ is the mass of the lens and $D_{\rm ol}$ and $D_{\rm os}$ represent 
the distances from the observer to the lens and source, respectively.

The light curve of a caustic-crossing binary lens event with a source 
composed of also binaries, which is the target event for followup monitoring 
to detect faint binary source companions by using the proposed method, 
deviates from the standard form in eq.\ (1) due to various effects.  These 
include the effects of lens and source binarity, and the finite source 
effect.  In the following subsections, we describe the basics 
of these effects.

\subsection{Binary Lens Events}
If an event is caused by a binary lens system, the locations of the 
individual images are obtained by solving the lens equation expressed in 
complex notations by
\begin{equation}
\zeta=z+\frac{m_1}{\bar{z}_1-\bar{z}}+\frac{m_2}{\bar{z}_2-\bar{z}},
\end{equation}
where $m_1$ and $m_2$ are the mass fractions of the individual lenses, $z_1$ 
and $z_2$ are the positions of the lenses, $\zeta=\xi+i\eta$ and $z=x+iy$ 
are the positions of the source and images, and $\bar{z}$ denotes the 
complex conjugate of $z$ (Witt 1990).  Note that all these lengths are 
dimensionless because they are normalized by the combined Einstein ring 
radius, which is equivalent to the Einstein ring radius of a single lens 
with a mass equal to the total mass of the binary.  Depending on the 
source position with respect to the lenses, there exist 3 or 5 solutions 
for the lens equation, and thus the same number of images.  The amplification 
of each image $A_i$  is given by the Jacobian of the lens equation evaluated 
at the image position, i.e.,
\begin{equation}
A_{0,i}=\left( \frac{1}{|{\rm det}\ J|} \right)_{z=z_i};\ \ 
    {\rm det}\ J=1-{\partial \zeta \over \partial \bar{z}}
     {\overline{\partial \zeta} \over \partial \bar{z}}.
\end{equation}
Then, the total amplification is obtained by summing the amplifications
of the individual images, i.e.\ $A_0=\sum A_{0,i}$.

The fundamental difference in the geometry of a binary lens system
from that of a single point-mass lens is the formation of caustics.  The 
caustic refers to the source position on which the amplification of a 
point source event becomes infinity, i.e., det $J=0$.  The set of 
caustics forms closed curves. The caustics takes various sizes and 
shapes depending on the projected separation $b$ (normalized by 
$\theta_{\rm E}$) and the mass ratio $q$ between the lens components.  
The probability of caustic crossing is maximized when the binary separation 
is equivalent to $\theta_{\rm E}$.  Whenever a source crosses the caustics, 
an extra pair of images appears (or disappears).  Hence the light curve of 
a caustic-crossing binary lens event is characterized by sharp spikes.  
Since the caustics form a closed curve, the source of a caustic-crossing 
event crosses the caustic at least twice.  Although the first caustic 
crossing is unlikely to be resolved due to its unpredictable and swift 
passage, it can be inferred from the enhanced amplification.  Then, by 
preparing intensive followup observations, resolving the second caustic 
crossing is possible.  From the surveys (EROS: Aubourg et al.\ 1993; 
MACHO: Alcock et al.\ 1993; OGLE: Udalski et al.\ 1993; DUO: Alard \& 
Guibert 1997) and followup observations (MPS: Rhie et al.\ 1999; PLANET: 
Albrow et al.\ 1998; MOA: Abe et al.\ 1997), numerous binary lens event 
candidates have been detected and a significant fraction of them are 
caustic-crossing events (Alcock et al.\ 2000).

\subsection{Finite Source Effect}
In most parts, the light curve of an observed caustic-crossing binary lens 
event is well described by the point-source approximation.  However, the 
part of the light curve near and during the caustic crossing deviates from 
this approximation.  This is because the source star in reality is not a 
point source and the gradient of the amplification near the caustics is 
very large.  As a result, different parts of the source star disk are 
amplified by considerably different amounts even with small differences 
in distance to the caustics (Schneider \& Weiss 1986; Witt \& Mao 1994).

The light curve of an extended source event is given by the intensity-weighted 
amplification averaged over the source star disk, i.e., 
\begin{equation}
A=\frac{\int^{2\pi}_0\int^{r_\star}_0 I(r,\theta)
A_0(\left\vert \rvec-\rvec_L \right\vert)
r\ dr\ d\theta}
{\int^{2\pi}_0\int^{r_\star}_0 I(r,\theta)r\ dr\ d\theta},
\end{equation}
where $r_\star$ is the radius of the source star, $I(r,\theta)$ is the 
surface intensity distribution of the source star, and the vectors $\rvec$ 
and $\rvec_L$ represent the displacement vector of the center of the source 
star with respect to the lens and the orientation vector of a point on the 
source star surface with respect to the center of the source star, 
respectively. Consequently, the amplification does not become infinite
even during the caustic crossing.

\subsection{Binary Source Events}
The light curve of an event with a source composed of double stars (binary 
source event) is very simple, as it is just the linear sum of the two light 
curves of the individual single source events (Griest \& Hu 1992).  Since 
the baseline flux is also the sum of the (unamplified) fluxes from the 
two source stars, the amplification becomes
\begin{equation}
A=\frac{A_1 F_{0,1}+A_2 F_{0,2}}{F_{0,1}+F_{0,2}},
\end{equation}
where $F_{0,i}$ and $A_i$ are the baseline fluxes and the amplifications of
the individual single source events and the subscripts $i=1$ and $2$ denote 
the primary and the companion source stars, respectively.

\section{Simulations}
The detectability of the light curve deviations induced by close-in binary 
source companions depends on various combinations of many parameters
such as those characterizing the binary source system (e.g., the projected 
separation and the relative phase between the source components) and the 
individual component source stars (e.g., the brightness and size).  
Consequently, it is difficult to present the detection efficiency as a 
function of a couple of parameters.  Therefore, we choose to evaluate the 
detectability by testing example light curves produced from the simulations 
of typical Galactic bulge events.  The simulations are performed in the 
following ways.

\begin{table}
\centering
\begin{minipage}{80mm}
\caption{The adopted values of the physical parameters of the source stars
for the tested example lensing events.  The values are adopted from
Allen (1973). }
\begin{tabular}{ccc}
\hline
\hline
\multicolumn{1}{c}{\ stellar\ } &
\multicolumn{1}{c}{\ absolute\ } &
\multicolumn{1}{c}{\ radius\ }\\
\multicolumn{1}{c}{ type} &
\multicolumn{1}{c}{\ magnitude\ } &
\multicolumn{1}{c}{($R_\odot$)}\\
\hline
Clump Giant   & +2.9 &  4.0 \\
G0 Dwarf      & +4.4 &  1.04\\
K0 Dwarf      & +5.9 &  0.85\\
M0 Dwarf      & +9.0 &  0.63\\
\hline
\end{tabular}
\end{minipage}
\end{table}

For the source stars, we test various combinations of stellar types, which 
well characterize the individual source components.  In Table 1, we list 
the source stellar types of the tested events along with the absolute 
magnitudes and radii, which are adopted from Allen (1973).  In the 
simulation, we assume that each source star has a uniform surface brightness 
distribution over its disk\footnote{If the source star disk is not uniform 
due to, for example, limb darkening (Witt 1995; Loeb \& Sasselov 1995; Gould 
\& Welch 1997; Valls-Gabaud 1998; Gaudi \& Gould 1999 ; Ignace \& Hendry 
1999) or spots (Heyrovsky \& Sasselov 2000; Han et al.\ 2000), the resulting 
light curve deviates from the uniform disk approximation.  However, we note 
that the amount of this additional deviation is slight.} and it is located 
at $D_{\rm os}=8$ kpc.  We also assume that the motion of each source is 
rectilinear. This is not only because the changes of the source positions 
caused by their orbital motion will be small during the caustic crossings, 
but also because what directly affects the detectability are the instantaneous 
projected separation, $\ell$ (normalized by $\theta_{\rm E}$), and the 
phase angle, $\alpha$, between the source components at the time of caustic 
crossing.

For the lens system, we assume that the mass ratio and the projected
separation between the lens components are $q=0.5$ and $b=1.2$.  The 
assumed Einstein ring radius crossing timescale (Einstein timescale) of 
the events is $t_{\rm E}=27$ days, which corresponds to the value of
a Galactic bulge event caused by a lens with a total mass and location of 
$\sim 0.3\ M_\odot$ and $D_{\rm ol}\sim 6$ kpc under the assumed lens-source 
transverse speed of $v\sim 150\ {\rm km}\ {\rm s}^{-1}$.

%Figure 1
\begin{figure}
\epsfysize=12cm
\centerline{\epsfbox{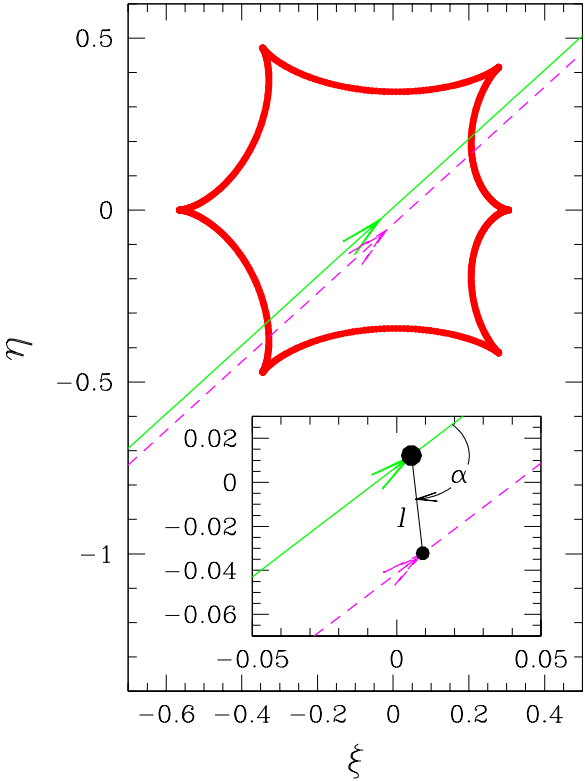}}
\caption{
The location of the caustics (the closed figure drawn by a thick solid
line) and the source star trajectories (straight lines) of one of the
tested caustic-crossing binary lens events with double source stars.
The coordinates are chosen so that the center of mass of the binary lens
is at the origin.  Both lenses are on the $\xi$ axis and the heavier lens
is to the right.  The separation and the mass ratio between the binary
lens components are $b=1.2$ and $q=0.5$.  The inset shows the locations
of the binary source stars and the phase angle at a particular moment.
}
\end{figure}

In Figure 1, we present the caustics of the lens system and the source 
trajectories of one of the tested example events.  We note that other 
tested events have similar trajectories except the variations caused by 
the differences in the values of $\ell$ and $\alpha$.

%Figure 2
\begin{figure}
\epsfysize=9cm
\centerline{\epsfbox{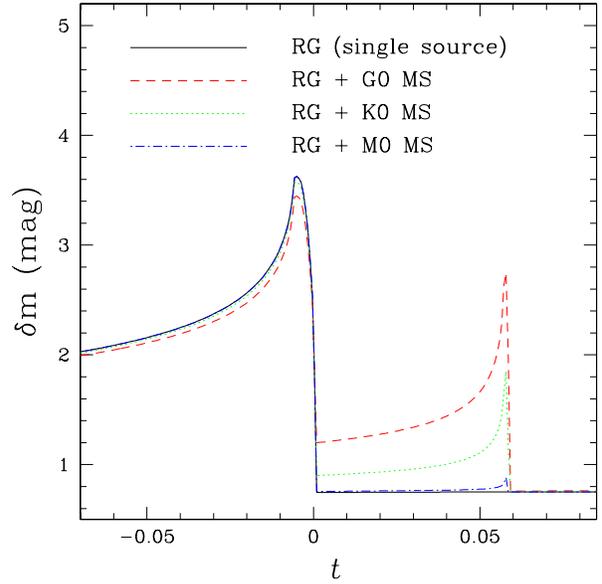}}
\caption{
The light curves of a caustic-crossing binary lens event with a source
also composed of binary stars where the primary is a red clump giant.
The curves with different line types represent those expected when the
primary source is accompanied by companions of G0, K0, and M0 main-sequence
stars, respectively.  The time is normalized by the Einstein time scale and
it is measured from the moment when the center of the primary star crosses
the caustic.  The source stars have a common separation and a phase angle
of $\ell=0.1$ and $\alpha=160^\circ$.
}
\end{figure}

%Figure 3
\begin{figure}
\epsfysize=9cm
\centerline{\epsfbox{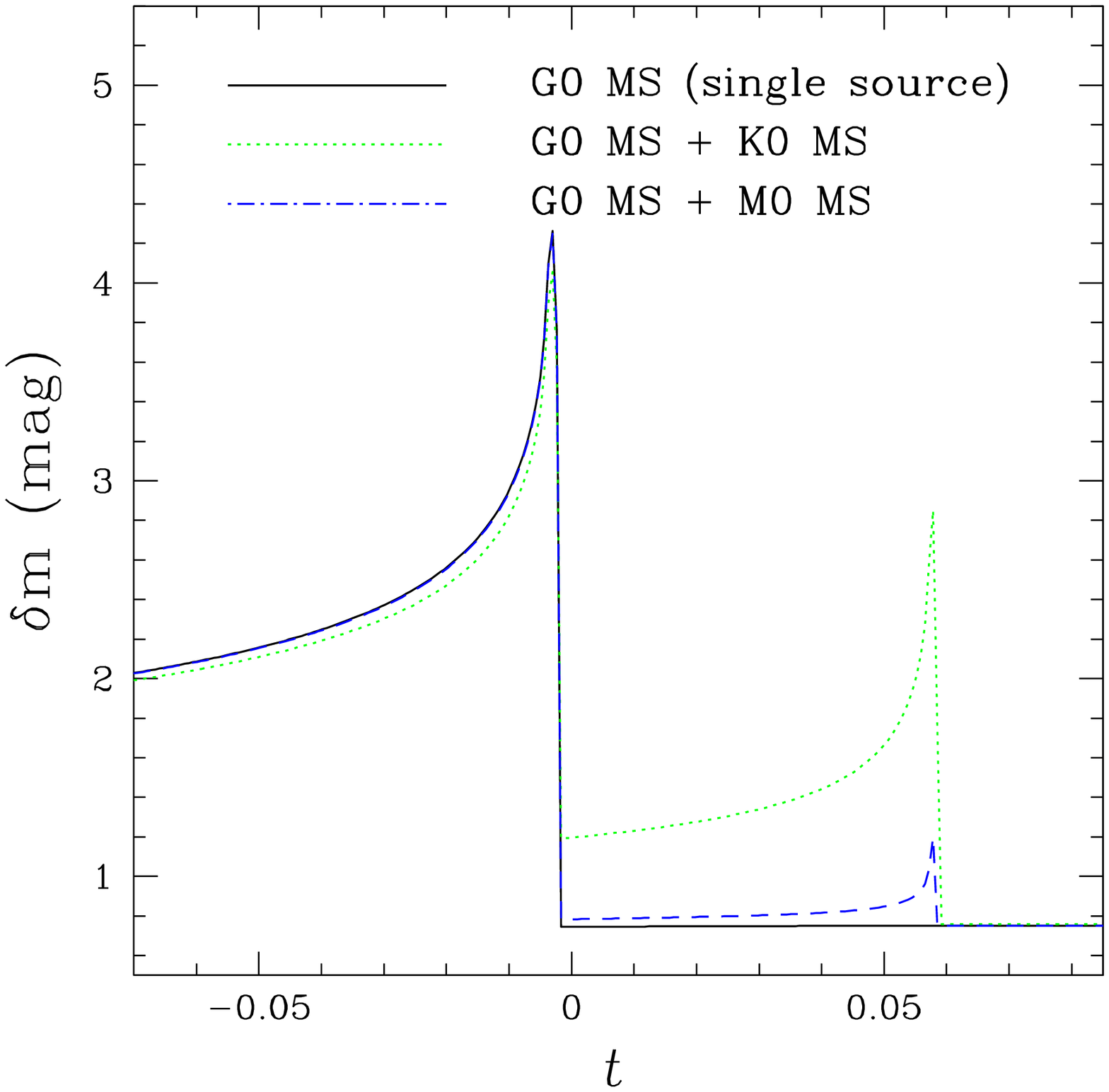}}
\caption{
Similar light curves as those in Fig.\ 2, but for events with different
combinations of source stellar types.  The individual light curves are
those expected when the primary of G0 main-sequence is accompanied by
companions of K0 and M0 main-sequence stars, respectively.  The source
stars have a common separation and a phase angle of $\ell = 0.1$ and
$\alpha=160^\circ$.
}
\end{figure}

In Figures 2 and 3, we present the light curves of events resulting from 
the various combinations of source stellar types.  We present only the part 
of the light curve around the second caustic crossing of the primary, 
because high time-resolution observation is possible at around this time.  
For the light curves in Fig.\ 2, the primary source star is a red clump 
giant (RG) and the corresponding companions are G0, K0, and M0 main-sequence 
(MS) stars, respectively.  For the light curves in Fig.\ 3, the primary is 
a G0 main-sequence and the companions are K0 and M0 main-sequence stars, 
respectively.  Note that the two tested stellar types of the primary source 
are the typical ones of Galactic bulge events.  For these events, we assume 
a common separation and a phase angle of $\ell=0.1$ and $\alpha=160^\circ$.  
Since the Einstein ring radius projected on the source plane of the example 
events is $\tilde{r}_{\rm E}=r_{\rm E}D_{\rm os}/D_{\rm ol} \sim 2.6$ AU, 
the assumed source separation corresponds to $\sim 0.25$ AU or $\sim 56\ 
R_\odot$.  We define $\alpha$ as the angle between the primary source 
trajectory and the line connecting the primary and secondary source stars 
measured in the clockwise sense (see the inset in Fig.\ 1). From the 
figures, one finds that the part of the light curve during the caustic 
crossing of the secondary source is spikier than the part during the 
primary source caustic crossing, implying that the secondary is more highly 
amplified.  This is because the secondary, in general, has a smaller size 
than the primary, and thus experiences less finite source effect.

%Figure 4
\begin{figure}
\epsfysize=9cm
\centerline{\epsfbox{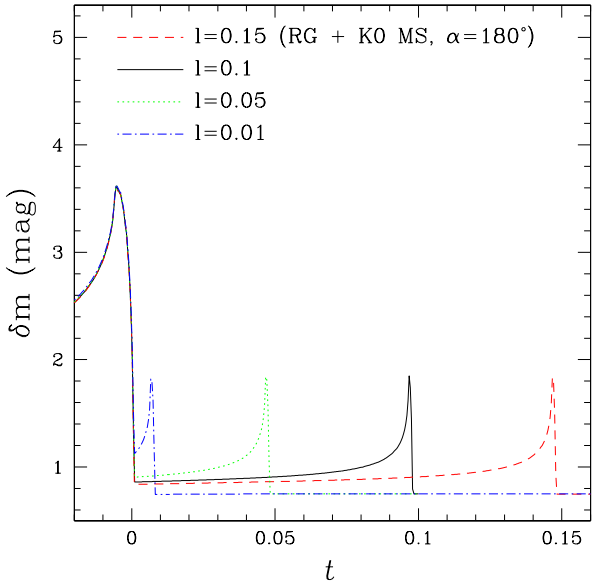}}
\caption{
Light curves of binary lens events occurred on binary source stars with
various projected separations.  The stellar types of the source stars are
a red clump giant for the primary and a K0 main sequence for the companion.
The assumed source orientation angle is $\alpha=180^\circ$.
}
\end{figure}

To investigate the dependency of the detectability on the binary source 
separation, we also perform simulations of events occurred on a binary 
source stars with various values of $\ell$, and the resulting light curves 
are presented in Figure 4.  For these events, the stellar types of the 
source stars are a red clump giant for the primary and a K0 main sequence 
for the companion, respectively.  We test four cases of source separation
of $\ell=$0.01, 0.05, 0.1, and 0.15.  To see the dependency of the 
detectability only on the separation, we assume $\alpha= 180^\circ$, 
implying that both stars are aligned and move along a common trajectory.

We also investigate the dependency of the detectability on the relative 
source phase by producing light curves of events occurred on source stars 
with various phase angles.  In Figure 5, we present the resulting light 
curves.  We test five cases of the phase angle of $\alpha=0^\circ$, 
$45^\circ$, $90^\circ$, $135^\circ$, and $180^\circ$.  For these events, 
we also assume a red clump giant primary and a K-type main sequence 
companion.  The source stars have a common separation of $\ell=0.1$.

%Figure 5
\begin{figure}
\epsfysize=9cm
\centerline{\epsfbox{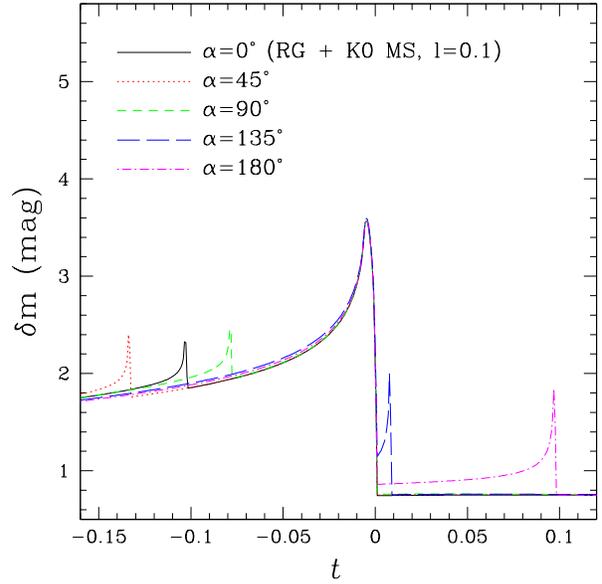}}
\caption{
Light curves of binary lens events occurred on binary source stars with
various phase angles.  The assumed stellar types of the primary and
companion are a red clump giant and a K-type main sequence, respectively.
The source stars have a common separation of $\ell=0.1$.
}
\end{figure}

\section{Detectability Evaluation}
In this section, we evaluate the detectability of the binary source 
induced deviations for the events produced by simulations in the previous 
section.

First, to be readily noticed, the light curve should deviate from that of 
a single source event with a significant fractional deviation.  The major 
part of the binary source event light curve is well approximated by that of 
a single source event with a baseline source flux equal to the sum of the 
fluxes from both source stars.  We, therefore, define the fractional excess 
amplification by
\begin{equation}
\epsilon = {A-A_{\rm s}\over A_{\rm s}},
\end{equation}
where $A$ and $A_{\rm s}$ represent the amplifications of the binary and 
single source events, respectively.  Then, we assume that the source 
companion can be detected if the fractional deviation is greater than a 
threshold value of $\epsilon_{\rm th}$.

\begin{table*}
\centering
\begin{minipage}{100mm}
\caption{The durations of deviations induced by the binary source companions
for the tested example lensing events.  For the definition of $t_{\rm dur}$,
see \S\ 4. }
\begin{tabular}{ccccc}
\hline \hline
\multicolumn{3}{c}{\ source stars\ } &
\multicolumn{1}{c}{$t_{\rm dur}$ } &
\multicolumn{1}{c}{related} \\
\multicolumn{1}{c}{stellar type} &
\multicolumn{1}{c}{separation ($\theta_{\rm E}$)} &
\multicolumn{1}{c}{phase angle} &
\multicolumn{1}{c}{(hrs)} &
\multicolumn{1}{c}{figure} \\
\hline
RG + G0 MS    & 0.1  & $160^\circ$ & 46.0 & Fig.\ 2\\
RG + K0 MS    & --   & --          & 32.4 & --     \\
\bigskip
RG + M0 MS    & --   & --          & 0.0  & --     \\
G0 MS + K0 MS & --   & --          & 46.0 & Fig.\ 3\\
\bigskip
G0 MS + M0 MS & --   & --          & 0.0  & --     \\
RG + K0 MS    & 0.01 & $180^\circ$ & 4.5  & Fig.\ 4\\
--            & 0.05 & --          & 28.5 & --     \\
--            & 0.1  & --          & 28.5 & --     \\
\bigskip
--            & 0.15 & --          & 28.5 & --     \\
RG + K0 MS    & 0.1  & $0^\circ$   & 4.5  & Fig.\ 5\\
--            & --   & $45^\circ$  & 7.8  & --       \\
--            & --   & $90^\circ$  & 5.8  & --    \\
--            & --   & $135^\circ$ & 5.2  & --       \\
--            & --   & $180^\circ$ & 28.5 & --       \\
\hline
\end{tabular}
\end{minipage}
\end{table*}

Second, even if the fractional deviation is substantial, the deviation 
cannot be detected if the absolute amount of the deviation is smaller than 
the photometric uncertainty.  Currently, microlensing followup observations 
are being (or planned to be) carried out by using the difference image 
analysis method for better photometric precision (D.\ Bennett, private 
communication).  This method measures the light variation by subtracting 
an observed image from a convolved and normalized reference image of the 
same field (Tomaney \& Crotts 1996; Alard \& Lupton 1998).  Then, the signal 
measured on the subtracted image is proportional to the source flux variation, 
i.e.\ $S\propto F_0(A-1)$, while the noise originates from both the source 
and background flux, i.e.\ $N\propto F_0A+B$, and thus the resulting 
signal-to-noise ratio is approximated by
\begin{equation}
S/N \sim F_0 (A-1) \left( {t_{\rm exp} \over F_0 A + 
	 \langle B \rangle }\right)^{1/2},
\end{equation}
where $F_0=F_{0,1}+F_{0,2}$ is the total baseline flux of the source stars, 
$t_{\rm exp}=2$ min is the adopted exposure time, and $\langle B \rangle$ 
is the average background flux within the point spread function (PSF).  
We assume that the observations are carried out in $I$ band\footnote{This 
is because in $I$ band more photons can be detected due to its broad band 
width and small amount of extinction.} by using a 1 m telescope with a 
CCD camera that detects $12\ e^{-}\ {\rm s}^{-1}$ for a star with $I=20$ 
mag.  Terndrup (1988) determined that the dereddened background surface 
brightness contributed by bulge stars is $\mu_I\sim 17.6\ {\rm mag}$ 
${\rm arcsec}^{-2}$.  With the adopted mean extinction of {\tt $\langle 
A_I\rangle=1.5$} towards the Galactic bulge field (Stanek 1996), and 
additional 30\% background flux from disk stars (Terndrup 1988) and 
sky background flux contribution of $\sim 1/4$ of stellar background, the 
average background flux within the PSF (with an angular area of $\Omega_{\rm 
psf}\sim 3.1\ {\rm arcsec}^{2}$) becomes $\langle B\rangle \sim 560\ e^{-1}
\ {\rm s}^{-1}$.  Once the signal-to-noise ratio is computed, the uncertainty 
of the light variation measurement in magnitude is computed by
\begin{equation}
\sigma = {\delta F/F\over 0.4\ {\rm ln}\ 10}
\sim
{1.09\over S/N},
\end{equation}
where $\delta F/F$ is the fractional uncertainty in the measured flux.
Then, if the deviation occurs when $A\sim 2$, the signal-to-noise ratio 
becomes {\tt $S/N\sim 14$} for the event with the primary source of the 
clump giant, and thus the photometric uncertainty is {\tt $\sigma\sim 0.08$} 
mag.  For the event with the primary source of the G0 main-sequence, the 
signal-to-noise ratio is {\tt $S/N\sim 3.7$} and the corresponding 
photometric uncertainty is {\tt $\sigma\sim 0.29$} mag.

The detection is additionally restricted by the duration of the deviations,
$t_{\rm dur}$.  We, therefore, estimate $t_{\rm dur}$ for the example events 
by defining $t_{\rm dur}$ as the length of time period during which the 
fractional and absolute deviations greater than $\epsilon_{\rm th}=0.1$ 
and $3\sigma$ level, respectively.  The resulting values of $t_{\rm dur}$ 
of the individual example events are listed in Table 2.  Considering the 
monitoring frequency of the current followup observations of $\sim 2\ 
{\rm hr}^{-1}$, we assume that the deviation can be detected if it 
continues for $t_{\rm dur}\geq 5\ {\rm hr}$.

The findings from the evaluation of the detectability are summarized as 
follows:
\begin{enumerate}
\item
For both red clump giant and G-type main-sequence primaries, we find 
that deviations induced by source companions brighter than K-type 
main-sequence or brighter can be efficiently detected by using the 
proposed method.  However, detecting deviations induced by M-type 
main-sequence companions will be difficult.

\item
In addition, we find that with the proposed method it will be possible to 
detect companions with separations as small as $\ell\sim 0.01$.  For a 
typical event with $\tilde{r}_{\rm E}\sim 2$ -- 3 AU, this separation
corresponds to $\sim 5\ R_\odot$.  Considering that the size of the a clump 
giant is $\sim 4\ R_\odot$, this implies that one can detect companions 
of nearly contact binary systems.\footnote{The light curve in such 
cases will be different from that presented in Fig.\ 4 due to the effect 
of stellar distortion.  However, analyzing the light curve variations caused 
by this effect is beyond the scope of this paper, and thus we use the 
approximation of circular source shape.} We also find that as long as the 
separation is larger than the limiting value, the dependency of the 
detectability on the source separation is weak.

\item
Unlike the weak dependency on $\ell$, we find that the detectability 
depends strongly on $\alpha$: it will be easier to detect the deviation 
if it occurs after the primary caustic crossing.  Although it will be 
possible to detect the deviation occurred before the primary caustic 
crossing, the duration of the deviation is significantly shorter than that 
of the deviation occurred after the primary caustic crossing.  This is 
because the amplification of the primary source flux at the time of 
deviation is high, and thus the fractional deviation caused by the 
companion is low.

\end{enumerate}

\section{Summary and Discussion}
In this paper, we propose a new method of detecting faint close-in stellar
companions by using microlensing.  The method utilizes a caustic-crossing 
binary lens event with a source also composed of binary stars, in which 
the companion is a faint star.  The detection is possible because the 
flux from the companion can be highly amplified when it crosses the lens 
caustic and is facilitated because the companion is likely to be more 
amplified than the primary due to the smaller size of the companion.  
From the simulations of Galactic bulge events under realistic conditions, 
we find that companions with K-type main-sequence or brighter of binary lens 
systems with separations down to $\sim ({\cal O}) 10^{-2}$ of the angular 
Einstein ring radius can be effectively detected by using the proposed 
method from the current type followup observations.

The proposed method of detecting source companions has both advantages and
disadvantages compared to that of detecting lens companions.  The most 
important disadvantage is that very faint or dark companions cannot be 
detected by using the proposed method, while the latter method can detect
companions regardless of their luminosities.  However, the proposed method 
has an important advantage of being able to detect very close companions 
with separations as small as $\sim ({\cal O})10^{-2}\theta_{\rm E}$.  By 
contrast, the efficiency of the latter method drops rapidly for binaries 
with separations $\lesssim 0.2\theta_{\rm E}$ (Gaudi \& Gould 1997).  
Therefore, the proposed method will make microlensing an important tool to 
detect binary companions by complementing the previous one.  In addition, 
since one can obtain the color and spectral information of the companion 
from followup multi-band photometric and spectroscopic observations 
performed during the deviation by taking advantage of its highly amplified 
flux, one can characterize the stellar type of the companion.
%Second, if multi-band photometric and spectroscopic 
%followup observations are conducted during the deviation induced by the 
%binary source companion, one can characterize the stellar type of the binary 
%source companion based on the spectral and color information.  This becomes 
%possible because most of the light during the deviation induced by the binary
%source companion directly comes from the companion itself if the deviation 
%occurs when the amplification of the primary source flux is not important.  
%With the known stellar type, the characteristics of the companion will be 
%better constrained than by the mass ratio $q$ (not the absolute mass) and the 
%instantaneous projected separation scaled by $\theta_{\rm E}$ (neither in 
%an absolute unit nor the true separation) obtained from the observation of 
%a binary lens event.

We would like to thank to P.\ D.\ Sackett for making useful comments about
the paper.  This work was supported by the grant (20006-113-03-2) of the 
Korea Science \& Engineering Foundation (KOSEF).

\clearpage

\end{document}